\documentclass[12pt]{article}
\usepackage{a4}
\usepackage{amsfonts}
\usepackage{amssymb}
\usepackage{lscape}
\usepackage{cite}
\usepackage{epsfig}
\usepackage{multirow}
\usepackage[all]{xy}
\usepackage{amsmath}
\author{}

\newcommand{\drawsquare}[2]{\hbox{%
\rule{#2pt}{#1pt}\hskip-#2pt
\rule{#1pt}{#2pt}\hskip-#1pt
\rule[#1pt]{#1pt}{#2pt}}\rule[#1pt]{#2pt}{#2pt}\hskip-#2pt
\rule{#2pt}{#1pt}}
\newcommand{\Ysymm}{\raisebox{-.5pt}{\drawsquare{6.5}{0.4}}\hskip-0.4pt%
         \raisebox{-.5pt}{\drawsquare{6.5}{0.4}}}
\newcommand{\Yasymm}{\raisebox{-3.5pt}{\drawsquare{6.5}{0.4}}\hskip-6.9pt%
        \raisebox{3pt}{\drawsquare{6.5}{0.4}}}

\newcommand{\be}{\begin{equation}}
\newcommand{\ee}{\end{equation}}
\newcommand{\ba}{\begin{array}}
\newcommand{\ea}{\end{array}}
\newcommand{\bea}{\begin{eqnarray}}
\newcommand{\eea}{\end{eqnarray}}

\newcommand{\ov}{\overline}
\def\IR{\relax{\rm I\kern-.18em R}}

\def\IP{\relax{\rm I\kern-.18em P}}
\def\inbar{\vrule height1.5ex width.4pt depth0pt}
\def\IC{\relax\,\hbox{$\inbar\kern-.3em{\rm C}$}}

\def\K3{{\bf K3}}

\def\ov{\overline}

\def\n2d{\cN_{V^*}^{\otimes 2}}

\def\IC{\mathbb{C}}

\def\IR{\mathbb{R}}

\def\IP{\mathbb{P}}

\def\cN{{\mathcal N}}

\begin{document}
\title{
\begin{flushright} \vspace{-2cm}
{\small UPR-1211-T\\
ROM2F/2009/16\\
} \end{flushright} \vspace{1.5cm}
Mass Hierarchies from MSSM Orientifold Compactifications }
 \vspace{1.0cm}
\author{\small  Mirjam Cveti{\v c}$^1$, James Halverson$^1$,  Robert Richter$^{1,2}$}

\date{}

\maketitle
\begin{center}
\emph{${}^1$ Department of Physics and Astronomy, University of Pennsylvania, \\
     Philadelphia, PA 19104-6396, USA }\\
\emph{${}^2$ Dipartimento di Fisica and Sezione I.N.F.N. \\ Universit\`a di Roma ``Tor Vergata''\\
Via della Ricerca Scientifica, 00133 Roma, Italy}\\
\vspace{0.5cm}
\tt{cvetic@cvetic.hep.upenn.edu, jhal@physics.upenn.edu, rrichter@roma2.infn.it}
\vspace{0.2cm}
\end{center}
\vspace{0.2cm}
\begin{abstract}

We investigate phenomenologically viable four- and five-stack MSSM D-brane quivers which
exhibit realistic fermion mass hierarchies. In our analysis, the mass hierarchies arise either from higher order terms containing the VEV's of SM singlets 
or from D-instanton effects, where the latter utilizes either family splitting or a 
factorizable Yukawa matrix. Extending the systematic bottom-up analysis of \cite{Cvetic:2009yh}, 
we present the only four-stack quiver with a semi-realistic Yukawa texture. Investigation
of five-stack MSSM models reveals many more quivers with analogous Yukawa textures, as well as a few
examples which exhibit three different mass scales for the up-quarks, down-quarks, and electrons.
Potential problems in this class of quivers are the presence of $U(1)$ instantons, which might lead to undesired effects, such as R-parity violating couplings, and the presence of dimension $5$ operators that could lead to rapid proton decay. We present a five-stack setup which overcomes all of these problems and exhibits three different mass scales for the up-quarks, down-quarks and electrons.

\end{abstract}

\thispagestyle{empty} \clearpage

\section{Introduction \label{sec intro}}

One of the most challenging problems in string theory is the quest for string vacua which give rise to the phenomenology observed in nature. Intersecting brane worlds (and their T-dual pictures) provide a very promising framework for such a quest \cite{Blumenhagen:2005mu, Marchesano:2007de,Blumenhagen:2006ci}.
In these models, the gauge groups appear on stacks of D-branes, which fill out four-dimensional space-time and wrap three-cycles in the internal manifold.  
Chiral matter is localized at the intersection of different stacks of D6-branes, and their multiplicity is given by the topological intersection number of the respective three-cycles that the D6-branes wrap.

Over the last decade, many supersymmetric string vacua have been constructed which give rise to an MSSM-like or GUT-like spectrum\footnote{For
original work on globally consistent non-supersymmetric intersecting D-branes, see
\cite{Blumenhagen:2000wh,Aldazabal:2000dg,Aldazabal:2000cn,Blumenhagen:2001te},
and for chiral globally consistent  supersymmetric ones, see
\cite{Cvetic:2001tj,Cvetic:2001nr}. For supersymmetric MSSM realizations, see \cite{Honecker:2004kb,Gmeiner:2007zz,Gmeiner:2008xq},
and for supersymmetric constructions within type II RCFT's, see \cite{Dijkstra:2004cc,Anastasopoulos:2006da}. First local (bottom-up) constructions were discussed in \cite{Antoniadis:2000ena,Aldazabal:2000sa}}. 
Given these, one can then investigate finer details of the MSSM, such as the hierarchies and textures of the Yukawa couplings\footnote{For such a study within  globally consistent supersymmetric three-family  MSSM-like constructions \cite{Cvetic:2001tj,Cvetic:2001nr}, see \cite{Cvetic:2002wh}.}.
In intersecting brane models, the Yukawa couplings can be calculated via CFT techniques, including the quantum prefactor \cite{Cvetic:2003ch}. They are suppressed by world-sheet instantons \cite{Aldazabal:2000cn, Cvetic:2001tj, Cremades:2003qj,Cvetic:2003ch}, which can lead to interesting hierarchies and may, in principle, reproduce the observed mass scales.  However, for a generic orientifold compactification, all world-sheet instanton contributions are expected to be of the same order and much fine-tuning is required to obtain the large hierarchies observed in nature.

Moreover, orientifold compactifications generically exhibit additional global $U(1)$'s, which are conserved at the perturbative level and thus forbid various Yukawa couplings. Such perturbatively forbidden couplings can be generated either by higher order couplings containing the VEV's of standard model (SM) singlets $\phi^i$ \cite{Dutta:2005bb,Chen:2008rx,Anastasopoulos:2009mr} or by 
D-instantons, which in type IIA are the so called E2-instantons \cite{Blumenhagen:2006xt,Ibanez:2006da,Florea:2006si}\footnote{For a recent review on the D-instanton effects see \cite{Blumenhagen:2009qh} and also \cite{Akerblom:2007nh,Cvetic:2007sj,Bianchi:2009ij}.}. The violation of the global $U(1)$'s in the Yukawa coupling is compensated for by the SM singlets or D-instantons. In both cases, the resulting coupling is suppressed:  For the generation via higher order couplings the suppression factor is $\prod_i \frac{\langle \phi^i\rangle}{M_s}$, where $M_s$ denotes the string mass and  $\langle \phi^i\rangle$ the VEV of the SM singlet $\phi^i$,  while for the non-perturbative generation the suppression factor is given by $e^{-S^{cl}_{E2}}$, where $S^{cl}_{E2}$ is the classical action of the E2-instantons. Note that these two mechanisms  give  hierarchies between perturbatively realized and perturbatively forbidden couplings, and more generally they provide hierarchies between Yukawa couplings which carry different global $U(1)$ charges. 

In \cite{Anastasopoulos:2009mr} (see also \cite{Kiritsis:2008ry}) the authors proposed that constructions where different families arise from different sectors provide a framework to explain the  mass hierarchies observed in nature. However, a systematic bottom-up analysis of three- and four-stack configurations which exhibit the exact MSSM spectrum\cite{Cvetic:2009yh} showed that the realization of such a D-brane spectrum often leads to phenomenological drawbacks, such as R-parity violating couplings, a $\mu$-term which is too large (see also \cite{Ibanez:2008my}), or too much family mixing.

Here we present an additional mechanism which naturally induces mass hierarchies within orientifold compactifications. This mechanism relies on the fact that non-perturbative effects often induce factorizable Yukawa matrices, which only generate masses for one family. The generation of masses for all three families then requires  two additional instantons, which generically give rise to different suppression factors and thus to different masses for each family. 

In this paper we investigate which MSSM D-brane quivers potentially exhibit observed mass hierarchies without inducing any undesired phenomenological drawbacks. In particular,  we extend a systematic bottom-up analysis of three- and four-stack MSSM quivers\cite{Cvetic:2009yh} by also requiring that the D-brane quivers can naturally account for the different mass scales of the MSSM, which are induced via the two mechanisms discussed above. We find that only one four-stack quiver allows for a semi-realistic Yukawa texture. This specific example, however,  predicts only  two different mass scales for the up-quarks.  After allowing for an additional $U(1)$ brane, i.e. a five-stack MSSM quiver,  we find many more setups which have Yukawa textures similar to the one in that specific four-stack quiver. Furthermore,  we  find a few setups which exhibit three different mass scales for the up-quarks, down-quarks, and electrons. In section \ref{sec five stack U(1) instanton} we display a five-stack quiver which gives rise to a semi-realistic Yukawa texture similar to the one encountered for the four-stack quiver. However, this setup exhibits an  U(1) instanton which generates an R-parity violating coupling.

A large number of quivers with realistic Yukawa textures exhibit quartic couplings, perturbatively or non-perturbatively realized, which might lead to a dangerously high proton decay rate. Both the four-stack quiver and the five-stack quiver, discussed in chapter \ref{sec four-stack} and section \ref{sec five stack U(1) instanton}, respectively, suffer from this drawback. In section \ref{sec perfect hierarchy}, we present one of the five-stack quivers which overcomes the issue of rapid proton decay, exhibits three different mass scales for the up-quarks, down-quarks and electrons, and explains the hierarchy between the $t$-quark mass and the masses of all other MSSM matter fields.  Thus this quiver passes all the stringent bounds arising from experimental observations and therefore provides an excellent starting point for further phenomenological investigation.

This paper is organized as follows. In chapter \ref{sec mass hierarchies}, we present in detail the two mechanisms which generate mass hierarchies. In chapter \ref{sec four-stack}, we present the only four-stack quiver which gives rise to semi-realistic Yukawa textures without inducing other phenomenological drawbacks. Since no four-stack quiver allows for three different mass scales for the up-quarks, down-quarks and electrons, we allow for an additional $U(1)$ brane in chapter \ref{sec five stack} and analyze a few representative examples with two and three different mass scales for the families.

\section{Mass Hierarchies in Orientifolds \label{sec mass hierarchies}}
In this section, we present two mechanisms which give rise to mass hierarchies in orientifold compactifications,
both of which are independent of any geometrical specifics of the compactification manifold. These compactifications generically exhibit multiple global $U(1)$'s, which are remnants of the generalized Green-Schwarz mechanism. Perturbative couplings must obey selection rules associated with these global $U(1)$'s, thus forbidding various couplings on the perturbative level.

Desired couplings which violate these selection rules can be induced by D-instanton effects \cite{Blumenhagen:2006xt, Ibanez:2006da, Florea:2006si} or by higher order couplings \cite{Dutta:2005bb,Chen:2008rx,Anastasopoulos:2009mr}. In the latter case, SM singlets $\phi^i$ compensate for the global $U(1)$ charges carried by the desired, but perturbatively forbidden, superpotential coupling. These singlets must acquire a VEV in order to induce the desired Yukawa coupling. The coupling is then suppressed by the factor $\prod_{i} \,\frac{\langle\, \phi^i\, \rangle }{M_s}$, where $M_s$ denotes the string mass and $i$ runs over the total number of singlets required to compensate for the global $U(1)$ charges. On the other hand, a D-instanton which carries the right global $U(1)$ charge can also generate the perturbatively missing coupling. In that case, the coupling is suppressed by the classical action of the instanton, which is related to the volume of the cycle wrapped by the instanton in the internal manifold.

Thus, in both cases, one naturally obtains a hierarchy between perturbatively realized couplings and couplings which are either induced non-perturbatively or by higher order couplings. This mechanism can be generalized to also achieve hierarchies between two couplings which are both perturbatively absent, and has been used to explain the three different mass scales for the three families of the MSSM \cite{Anastasopoulos:2009mr}. Let us briefly review that mechanism in more detail.
\subsection{Hierarchy from Family Splitting \label{sec Family splitting}}
Depending on the choice of hypercharge, various MSSM matter fields might have different origins. For instance, in the Madrid embedding\footnote{The Madrid quiver is based on four stacks of D-branes $a$, $b$, $c$ and $d$, which give rise to the gauge symmetry $U(3)_a \times U(2)_b \times U(1)_c \times U(1)_d$. Applying the Green-Schwarz mechanism gives the standard model gauge symmetry $SU(3)_a \times SU(2)_b \times U(1)_Y$ in four dimensions, where the hypercharge is given by the linear combination \eqref{eq: hypercharge madrid quiver} of the anomalous $U(1)$'s.}
\begin{align}
U(1)_Y=\frac{1}{6} U(1)_a +\frac{1}{2} U(1)_c +\frac{1}{2} U(1)_d,
\label{eq: hypercharge madrid quiver}
 \end{align}
the left-handed quarks $q_L$ can transform as $(a,b)$ or $(a,\ov b)$, and the right-handed down-quarks $d_R$ as $\Yasymm_a$, $(\ov a, c)$ or $(\ov a ,d)$ \footnote{Here $a$ ($\ov a$) stands for the fundamental $3$  (anti-fundamental $\ov 3$) of $SU(3)$ and analogously for the other D-brane gauge groups. Moreover, we choose the convention that positive intersection number $I_{ab}$ between the stacks $a$ and $b$ corresponds to a transformation behavior $(a, \ov b)$. \label{footnote intersection}}. Assuming that every coupling $q^I_L \, H_d \, d^J_R$ is realized either perturbatively, via higher order couplings, or non-perturbatively, one can naturally obtain hierarchies for the masses. In \cite{Anastasopoulos:2009mr} the authors analyzed which Yukawa matrix textures can lead to three different mass scales.
For the quark Yukawa matrices, the potential matrices take the form
\begin{align}
Y^1=\left(
\begin{array}{ccc}
A &  B &  B \\
C & D& D \\
C & D & D
\end{array}
\right) \,\,\,\,\,\,\,\,\,
Y^2=\left(
\begin{array}{ccc}
A &  B &  C \\
A & B& C  \\
A & B & C
\end{array}
\right)
\,\,\,\,\,\,\,\,\,
Y^3=\left(
\begin{array}{ccc}
A &  B &  C \\
D & E& F  \\
D & E & F
\end{array}
\right)\,\,,
\label{eq: Yukawa matrices quarks}
\end{align}
where same letters denote terms of the same order. The Yukawa matrix $Y^1$ arises when one left-handed quark $q_L$ transforms as $(a,b)$, while the other two transform as $(a, \ov b)$, and when the right-handed quarks split in such a way that exactly two come from one sector and the third from a different one. The Yukawa matrix $Y^2$ is realized when all the $q_L$'s are localized at the same sector, but there is a maximal distribution of the right-handed quarks, i.e. all three right-handed quarks arise from different sectors. Finally, the quark Yukawa matrix takes the form $Y^3$ when the three $q_L$'s arise from two different origins and there is  a maximal splitting in the right-handed quarks. Note that the latter matrix might still have three different mass scales even if some of the six parameters are zero. For the matrix $Y^2$ all entries $A$, $B$ and $C$ have to be non-vanishing, while for $Y^1$ the entry $A$ could be zero and one still gets three different mass scales.

For the leptonic coupling $L^I \, H_d\, E^J_R$ there is an additional Yukawa matrix which can account for the three mass scales,
\begin{align*}
Y^4=\left(
\begin{array}{ccc}
A &  B &  C \\
D & E& F  \\
G & H & I
\end{array}
\right)\,\,.
\end{align*}
Such a texture arises when all of the $L^I$ and $E^I_R$, respectively, come from different sectors.
Let us emphasize that the different letters in the four Yukawa matrices are generically of different order. 

Note that for this mechanism, there is no crucial difference whether the entries are induced non-perturbatively or via higher order couplings containing the VEV's of SM singlets, where different singlets generically acquire VEV's of different scales.

\subsection{Hierarchy from Factorizable Yukawa Matrices \label{sec factorizable Yukawa matrices}}
While the previous mechanism does not rely on the presence of non-perturbative effects, the mechanism presented below is purely non-perturbative \cite{Blumenhagen:2007zk,Ibanez:2008my,Cvetic:2009yh}. For the sake of concreteness, consider a setup where three $U(1)$ branes $a$, $b$ and $c$ exhibit the intersection pattern\footnote{As already mentioned in footnote \ref{footnote intersection}, positive intersection number $I_{ab}=K$ corresponds to $K$ fields transforming as $(a,\ov b)$.}
\begin{align*}
I_{ab}=K \qquad I_{ac}=0 \qquad I_{bc}=K\,\,,
\end{align*}
where we denote fields arising from the $ab$ sector as $\Phi^I$ and fields arising from the $bc$ sector as $\widetilde \Phi^I$.  The superpotential term
\begin{align*}
M_s \,\, \Phi^I_{(1,-1,0)} \, \widetilde \Phi^J_{(0,1,-1)}
\end{align*}
is perturbatively forbidden, where the subscript denotes the charge under the global $U(1)_a$, $U(1)_b$ and $U(1)_c$, respectively. An instanton with the intersection pattern 
\begin{align*}
I_{E2a}=1 \qquad I_{E2b}=0 \qquad I_{E2c}=-1
\end{align*}
carries the global $U(1)$ charges  $Q_{E2}(a) = -1$, $Q_{E2}(b)=0$ and $Q_{E2}(c)=1$, and therefore can induce the perturbatively missing coupling. Such an instanton exhibits two charged zero modes, $\ov \lambda_a$ and  $\lambda_c$, and its action
generically takes the form
\begin{align}
S_{E2}= S^{cl}_{E2} + Y^{IJ} \,\ov \lambda_a \,\Phi^I\,\widetilde \Phi^J \,  \lambda_c   \label{eq: instanton action non-factorized} \,\,,
\end{align}
where the prefactor $Y^{IJ}$ contains the world-sheet instanton contributions and is expected to be of the same order for all combinations  $IJ$.
Performing the path integral
\begin{align*}
\int d^4 x \, d^2 \theta\, d \ov \lambda_a \, d \lambda_c \,\,e^{-S_{E2}}
\end{align*}
gives the result
\begin{align*}
M^{IJ}= Y^{IJ}\,e^{-S^{cl}_{E2}}\, M_s
\end{align*}
and leads generically to a mass matrix whose $K$ eigenvalues are all of the same order.

On the other hand an instanton with the intersection pattern
\begin{align}
I_{E2a}=1 \qquad I_{E2b}=0 \qquad I_{E2c}=-1 \qquad I^{{\cal N}=2}_{E2b}=1 \label{eq: instanon factorizable intersection pattern}
\end{align}
also carries total global $U(1)$ charge $Q_{E2}(a)=-1$, $Q_{E2}(b)=0$ and $Q_{E2}(c)=1$, but exhibits two additional charged zero modes
$\lambda_b$ and $\ov \lambda_b$. The generic instanton action has additional terms compared to \eqref{eq: instanton action non-factorized}, 
\begin{align*}
S_{E2}= S^{cl}_{E2} + Y^{IJ} \,\ov \lambda_a \,\Phi^I\,\widetilde \Phi^J \,\lambda_c + Y^{I} \,\ov \lambda_{a} \, \Phi^{I}   \lambda_b + Y^{J}\, \ov \lambda_{b} \, \widetilde \Phi^{J} \lambda_c\,\,.
\end{align*}
Now the measure in the path-integral
\begin{align*}
\int d^4 x \, d^2 \theta\, d \ov \lambda_a\,  d \lambda_b\, d \ov \lambda_b \, d \lambda_c \,\,e^{-S_{E2}}
\end{align*}
contains the additional Grassmann variables $\lambda_b$ and $\ov \lambda_b$, which must be saturated in order to get a non-vanishing answer. The resulting instanton induced mass term is given by
\begin{align}
M^{IJ}= Y^{I} \, Y^{J} \,e^{-S^{cl}_{E2}}\, M_s\,\,.
\end{align}
This matrix factorizes and thus only one linear combination of $\Phi^I\, \widetilde \Phi ^J$ becomes massive. In order to give masses to all $K$ families, an additional $K-1$ instantons of the type \eqref{eq: instanon factorizable intersection pattern} are required. The resulting mass eigenvalues depend on the suppression factor of these additional instantons, which generically are all different from each other, since the instantons wrap different cycles in the internal manifold. Thus, if non-perturbative effects induce a factorizable Yukawa matrix, one naturally obtains different mass scales for different families.

In the following chapters, we will make use of both mechanisms to explain the observed mass hierarchies in nature.

\section{Four-stack Quivers \label{sec four-stack}}
In \cite{Cvetic:2009yh}, the authors systematically analyzed different four-stack quivers which give rise to the exact MSSM spectrum. While on the order of 10000 setups satisfy the string consistency conditions, namely tadpole cancellation and the conditions due to the presence of a massless hypercharge $U(1)_Y$, only a small subset fulfill the bottom-up constraints. The latter include the presence of all the MSSM Yukawa couplings, a  $\mu$-term of the correct order, and the absence of R-parity violating couplings on the perturbative and non-perturbative level. Additionally, all setups are required to account for the observed CKM matrix and exhibit a mechanism which explains the smallness of the neutrino masses. Finally, to account for the large top mass, the authors required the presence of the $t$-quark Yukawa coupling on the perturbative level\footnote{The authors also allowed for the possibility that the $t$-quark Yukawa coupling can be realized non-perturbatively. In that case, the couplings $q_L H_d d_R$ and $L H_d E_R$ must also be induced non-perturbatively, where the suppression factor for the latter two couplings is larger than for the $t$-quark Yukawa coupling.}. All the top-down and bottom-up constraints implemented in the systematic analysis of  \cite{Cvetic:2009yh} are briefly summarized in appendix \ref{app top-down bottom-up}.

One of the results of this four-stack analysis was that different origins for different families, which could be an explanation for the hierarchies, are phenomenologically not favored. This is basically due to the fact that in such configurations the instantons required to induce the perturbatively absent, but desired, MSSM couplings also induce R-parity violating couplings or a $\mu$-term which is too large (see also \cite{Ibanez:2008my})\footnote{Though they could also generate the missing couplings via higher order couplings, SM singlets are not of any help, since generically they would also induce a too large $\mu$-term or R-parity violating couplings.}.

The masses of the MSSM indicate not only hierarchies between different families, but also a hierarchy between the $t$-quark mass and the masses of all other MSSM matter fields. Such hierarchy is naturally achieved when only the $t$-quark Yukawa coupling is perturbatively realized, while all other Yukawa couplings are forbidden due to violation of global $U(1)$'s. In this chapter, we present a four-stack quiver which realizes the latter and thus explains the hierarchy between the $t$-quark mass and all other masses. Moreover, this quiver exhibits hierarchies between different families based on the two mechanisms discussed in the previous chapter: family splitting and non-perturbative effects which induce factorizable Yukawa matrices.

As already suggested in \cite{Anastasopoulos:2009mr}, the Madrid embedding 
\begin{align*}
U(1)_Y= \frac{1}{6} U(1)_a + \frac{1}{2} U(1)_c +\frac{1}{2} U(1)_d
\end{align*}
 is the most promising hypercharge embedding for realizing mass hierarchies, since the matter fields have many potential origins. Below we summarize the potential origins of all the MSSM matter fields for the Madrid embedding
\begin{align*}
q_L\,:&\qquad (a,\ov{b}),\,\,\,\,(a,b)\\
u_R\,:&\qquad (\ov{a},\ov{c}),\,\,\,\,(\ov{a},\ov d)\\
d_R\,:&\qquad {\Yasymm}_a,\,\,\,\,(\ov{a},c),\,\,\,\,(\ov{a},d)\\
L\,\,:&\qquad (b,\ov{c}),\,\,\,\,(\ov{b},\ov{c}),\,\,\,\,(b,\ov d),\,\,\,\,(\ov b,\ov d)\\
E_R\,:&\qquad (c,d),\,\,\,\,{\Ysymm}_c,\,\,\,\,{\Ysymm}_d\\
N_R\,:&\qquad {\Yasymm}_b,\,\,\,\,\ov{\Yasymm}_b,\,\,\,\,(c, \ov d),\,\,\,\, (\ov c, d) \\
H_u\,:&\qquad (\ov b,c),\,\,\,\,(b,c),\,\,\,\,(\ov b,d),\,\,\,\,(b,d)\\
H_d\,:&\qquad (b,\ov{c}),\,\,\,\,(\ov{b},\ov{c}),\,\,\,\,(b,\ov d),\,\,\,\,(\ov b,\ov d)\,\,\,.\\
\end{align*}

In addition to the top-down and bottom-up constraints laid out in appendix \ref{app top-down bottom-up} (see also \cite{Cvetic:2009yh}), we impose the additional condition that the D-brane quiver exhibits mass scales similar to the ones observed in nature. Here we require that the hierarchies are generated via the two mechanisms presented in chapter \ref{sec mass hierarchies}. We find only one D-brane quiver which fulfills all the top-down and bottom-up constraints while also exhibiting a semi-realistic Yukawa structure. Subsequently, we will analyze this D-brane quiver in detail with respect to its Yukawa couplings and textures.  

In Table \ref{table spectrum four stack quiver} we display the origin and transformation behavior of the MSSM matter content.

\begin{table}[h] \centering
\begin{tabular}{|c|c|c|c|c|}
\hline
 Sector & Matter Fields &  Transformation & Multiplicity & Hypercharge\\
\hline \hline
 $ab$                            & $q_L$  & $(a,\overline{b})$ & $3$& $\frac{1}{6}$ \\
\hline
 $ac'$                            & $u_R$  & $(\overline{a},\overline{c})$  & $1 $ & $-\frac{2}{3}$ \\
\hline
 $ad'$                            & $u_R$  & $(\overline{a},\ov d)$  & $2 $ & $-\frac{2}{3}$ \\
\hline
$aa'$ & $d_R$ & ${\Yasymm}_a$ & $3$& $\frac{1}{3}$  \\
\hline
$bc$                            & $L$  & $(b,\overline{c})$  & $3$& $-\frac{1}{2}$ \\
\hline
$bc'$                            & $H_u+H_d$  & $(b,c)+(\overline{b},\overline{c})$ & $1$ & $\frac{1}{2} \,\,\, -\frac{1}{2} $  \\
\hline
$bb'$                            & $N_R$  & $\overline{{\Yasymm}}_b $  & $3$ & $0$ \\
\hline
$cd'$                            & $E_R$  & $(c, d)$  & $3 $ & $1$ \\
\hline
\end{tabular}
\caption{\small{Spectrum for the $4$-stack quiver with $U(1)_Y=\frac{1}{6}\,U(1)_a+\frac{1}{2}\,U(1)_c+ \frac{1}{2}\,U(1)_d$.}}
\label{table spectrum four stack quiver}
\end{table}\vspace{5pt}

The only perturbatively realized couplings are
\begin{align*}
&<{q^I_L}_{(1,-1,0,0)}\, {H_u}_{(0,1,1,0)}\, {u_R^1}_{(-1,0,-1,0)}>\qquad <L^I_{(0,1,-1,0)}\, {H_u}_{(0,1,1,0)} \, {N_R^J}_{(0,-2,0,0)}>
\\
& \qquad \qquad \qquad \qquad \qquad <{H_u}_{(0,1,1,0)}\, {H_d}_{(0,-1,-1,0)}>\,\,.
\end{align*}
The other couplings are expected to be induced by instantons\footnote{The perturbatively missing couplings can also be generated via higher order couplings, where the SM singlets acquire a VEV. As explained in chapter \ref{sec mass hierarchies}, the consequences of higher order couplings are analogous to those of instantons. From now on we assume that the perturbatively forbidden couplings are induced via D-instantons.}. Let us examine if the non-perturbative effects can give rise to mass hierarchies
observed in nature. We start by analyzing the up-quark coupling. The missing coupling
\begin{align*}
<{q^I_L}_{(1,-1,0,0)}\, {H_u}_{(0,1,1,0)}\, {u^{2,3}_R}_{(-1,0,0,-1)}>
\end{align*}
 is generated by an instanton with the intersection pattern
\begin{align*}
I_{E2_1a}=0 \qquad I_{E2_1b}=0 \qquad I_{E2_1c}=1 \qquad  I_{E2_1d}=-1 \,\,.
\end{align*}
Including the perturbatively realized coupling $q^I_{L} H_u u^1_R$, the resulting up-quark mass matrix takes the form
\begin{align}
M^{IJ}_{U}= \,\langle \, H_u \, \rangle   \,\,\left(
\begin{array}{ccc}
g^u_{11} & g^u_{12} \,e^{-S^{cl}_{E2_1}} & g^u_{13} \,e^{-S^{cl}_{E2_1}}\\
g^u_{21} & g^u_{22} \,e^{-S^{cl}_{E2_1}} & g^u_{23} \,e^{-S^{cl}_{E2_1}}\\
g^u_{31} & g^u_{32} \,e^{-S^{cl}_{E2_1}} & g^u_{33} \,e^{-S^{cl}_{E2_1}}\\
\end{array}
\right)\,\,,
\label{eq: up-quark matrix 4-stack}
\end{align}
where $g^u_{ij}$ are the world-sheet instanton contributions and $\langle \, H_u \, \rangle  $ denotes the VEV of the Higgs field, which is of the order $100\, GeV$. Note that for generic values of $g^u_{ij}$, the up-quark matrix \eqref{eq: up-quark matrix 4-stack} gives rise to two different scales for the mass eigenvalues
\begin{align*}
m_{t} : m_{c}: m_{u} \simeq 1 : e^{-S^{cl}_{E2_1}}: e^{-S^{cl}_{E2_1}}\,\,.
\end{align*}
This is not desired, of course, since we observe three different scales in nature. World-sheet instantons can, in principle, account for the hierarchy between the two lightest families, but that crucially depends on the internal geometry and generically requires some amount of fine-tuning.

Let us turn to the down-quark coupling
\begin{align*}
<{q^I_L}_{(1,-1,0,0)}\, {H_d}_{(0,-1,-1,0)}\, {d^J_R}_{(2,0,0,0)}>\,\,,
\end{align*}
which could be induced by an instanton, $E2_2$, with the intersection pattern
\begin{align}
I_{E2_2a}=1 \qquad I_{E2_2b}=-1 \qquad I_{E2_2c}=-1 \qquad  I_{E2_2d}=0 \,\,.
\label{eq: intersection pattern E2_2}
\end{align}
Generically, such an instanton exhibits the action
\begin{align*}
S_{E2_2}= S^{cl}_{E2_2} +  Y^I_{q_L}\, \ov \lambda_{a} \,q^I_L\, \lambda_{b} +  Y_{H_u}\,\lambda_{b} \,H_d\, \lambda_c + Y^J_{d_R}\ov \lambda_a \, d^J_R \, \ov \lambda_a\,\,,
\end{align*}
where the coefficients $Y$ contain the world-sheet instantons and can be computed using CFT techniques \cite{Cvetic:2007ku}.
Performing the path integral
\begin{align*}
\int d^4x\, d^2 \theta\, d^3 \ov \lambda_a\, d^2 \lambda_b \,d \lambda_c \,\, e^{-S_{E2_2}}
\end{align*}
gives the Yukawa coupling
\begin{align*}
Y^{IJ}_D \, e^{-S^{cl}_{E2_2}} \, q^I_L\,H_d\, d^J_R\,\,,
\end{align*}
where $Y^{IJ}_D$ contains the world-sheet instanton contributions $Y^I_{q_L}$, $Y_{H_u}$ and $Y^J_{d_R}$. Note that the Yukawa matrix factorizes
\begin{align}
Y^{IJ}_D\sim Y^I_{q_L} \, Y^J_{d_R}, \label{eq: d-quark factorizable}
\end{align}
and thus one needs three different instantons, $E2_2$, $E2'_2$ and $E2''_2$, with the intersection pattern \eqref{eq: intersection pattern E2_2}. The resulting down-quark mass matrix then takes the form
\begin{align}
M^{IJ}_{D}= \,\langle \, H_d \, \rangle   \,\,\left(
\begin{array}{ccc}
a_{11} & a_{12} & a_{13} \\
a_{21} & a_{22} & a_{23} \\
a_{31} & a_{32} & a_{33} \\
\end{array}
\right)\,\,,
\label{eq: d-quark matrix 4-stack}
\end{align}
with $\langle \, H_d \, \rangle $ on the order of $100\,  GeV$ and 
\begin{align*}
a_{IJ}=Y^{IJ}_D\,\, e^{-S^{cl}_{E2_2}}+Y'^{IJ}_D  \,\,e^{-S^{cl}_{E2'_2}} +Y''^{IJ}_D\,\,  e^{-S^{cl}_{E2''_2}}\,\,,
\end{align*}
where the $Y^{IJ}_D$, $Y'^{IJ}_D$ and $Y''^{IJ}_D$ factorize as in equation \eqref{eq: d-quark factorizable}. The down-quark matrix \eqref{eq: d-quark matrix 4-stack} generically exhibits three different mass eigenvalues, whose ratio is roughly given by
\begin{align*}
m_b : m_s : m_d  \simeq  e^{-S^{cl}_{E2_2}} :  e^{-S^{cl}_{E2'_2}}:  e^{-S^{cl}_{E2''_2}}\,\,.
\end{align*}
In order to obtain the observed hierarchy between the $t$-quark and $b$-quark masses, the suppression factor $e^{-S^{cl}_{E2_2}}$ is expected to be in the range $10^{-2}-5 \cdot 10^{-2}$.

Note that the coupling $q^I_L\, H_d \, d^J_R$ is an example where the hierarchies between different families are not due to family splitting, but are rather due to the fact that  an instanton with the intersection pattern \eqref{eq: intersection pattern E2_2} induces a factorizable Yukawa matrix. One needs three different instantons with this intersection pattern to generate masses for all three families. Each of those three instantons wraps a different cycle in the internal manifold and thus leads to a different suppression factor. By this mechanism, the quiver exhibits the observed hierarchy.

For the lepton Yukawa coupling
\begin{align*}
<L^{I}_{(0,1,-1,0)} \,{H_d}_{(0,-1,-1,0)}\,  {E^J_R}_{(0,0,1,1)}>\,\,,
\end{align*}
which is perturbatively forbidden, we encounter a very similar situation.
There are potentially two different classes of instantons which can generate this perturbatively missing coupling. Their intersection patterns are given by
\begin{align}
I_{E2_3a}=0 \qquad I_{E2_3b}=0 \qquad I_{E2_3c}=-1 \qquad I_{E2_3d}=1
\label{eq: intersection pattern E2_3}
\end{align}
and
\begin{align}
I_{\widetilde{E2}_3a}=0 \qquad I_{\widetilde{E2}_3b}=0 \qquad I_{\widetilde{E2}_3c}=-1 \qquad I_{\widetilde{E2}_3d}=1 \qquad I^{{\cal N}=2}_{\widetilde{E2}_3c}=1\,\,.
\label{eq: intersection pattern tilde E2_3}
\end{align}
An instanton of type \eqref{eq: intersection pattern E2_3} does not induce a factorizable Yukawa matrix. In that case, the electron masses are expected to be all of the same order, which is in contradiction to experimental observations. On the other hand, an instanton exhibiting the intersection pattern \eqref{eq: intersection pattern tilde E2_3} gives rise to a factorizable Yukawa matrix. In the following we show that factorization very explicitly. 

The instanton $\widetilde{E2}_3$ exhibits the zero modes 
 $\lambda^1_c$, $\lambda^2_c$, ${\ov \lambda}_c$ and ${\ov \lambda}_d$ and gives rise to the action
\begin{align*}
S_{\widetilde{E2}_3}&=  S^{cl}_{\widetilde{E2}_3}  +  Y^I_{E}\, \, \ov \lambda_c \,E^I_R \,{\ov \lambda}_d +\, Y^I_L \,\, \lambda^1_c \, L^I H_d\, \lambda^2_c + \,\widetilde{Y}^I_L \,\,\lambda^2_c \, L^I H_d\, \lambda^1_c \\& \qquad \qquad+ Y^{IJ}_{L\, H\, E} \,\, \lambda^1_c \, L^I H_d E^J_R \, {\ov \lambda}_d  + Y'^{IJ}_{L\, H\, E} \,\, \lambda^2_c\,  L^I H_d E^J_R \, {\ov \lambda}_d \,\,.
\end{align*}
Performing the path integral 
\begin{align*}
\int d^4 x \,d^2 \theta \,d^2 \lambda_c\, d \ov \lambda_c \,d  {\ov \lambda}_d \,\,e^{-S_{\widetilde{E2}_3}}
\end{align*}
then leads to the electron Yukawa matrix
\begin{align}
Y^{IJ}_{E}= e^{-S^{cl}_{\widetilde{E2}_3}}\,\,\langle \, H_d \, \rangle  \, \left( Y^{I}_{L} - \widetilde{Y}^{I}_{L} \right)\, Y^{J}_E \,\,,
\end{align}
which indeed factorizes. Thus one needs three different instantons, $\widetilde{E2}_3$, $\widetilde{E2}'_3$ and $\widetilde{E2}''_3$, each with intersection pattern \eqref{eq: intersection pattern tilde E2_3}, to induce masses for all three families\footnote{There is another possibility, where only two instantons have the intersection pattern \eqref{eq: intersection pattern tilde E2_3}, and one instanton has the intersection pattern \eqref{eq: intersection pattern E2_3}. In this case, the latter would have the largest suppression
factor.}. This could explain the observed hierarchies, due to their different suppression factors. To match experimental observations, we expect the ratios between the different instanton suppressions to satisfy
\begin{align*}
m_{\tau} : m_{\mu} :m_{e} \simeq e^{-S^{cl}_{\widetilde{E2}_3}}: e^{-S^{cl}_{\widetilde{E2}'_3}}: e^{-S^{cl}_{\widetilde{E2}''_3}}\,\,.
\end{align*}
To get the desired mass hierarchy between $t$-quark and the $\tau$ lepton, the suppression factor $e^{-S^{cl}_{\widetilde{E2}_3}}$ is expected to be of the order  $10^{-2}$.

As observed above, the Dirac neutrino masses are perturbatively realized for all three families and are therefore expected to be of the same order as the $t$-quark mass. The presence of a large Majorana mass term for the right-handed neutrinos would give a natural explanation for the smallness of the neutrino masses, via the seesaw mechanism. Such a Majorana mass term can be induced by an instanton with the intersection pattern
\begin{align}
I_{E2_4a}=0 \qquad I_{E2_4b}=-2 \qquad I_{E2_4c}=0 \qquad I_{E2_4d}=0\,\,.
\label{eq: intersection pattern E2_4}
\end{align}
Even though, at first sight, it seems that the Majorana mass matrix factorizes and thus would only generate a seesaw mass for one neutrino family, we show in appendix \ref{app neutrino mass} that the Majorana mass matrix for a generic compactification does not factorize. With a string mass of the order $M_s \simeq 10^{18} \, GeV$, we expect the suppression factor to be in the range $10^{-5}-10^{-2}$ in order to obtain seesaw neutrino masses in the observed range.

Finally, let us draw the attention to the following two dimension $5$ operators, 
\begin{align}
\frac{\kappa}{M_s}  \,\,q_L\, q_L \,q_L \,L \qquad \text{and} \qquad \frac{\kappa'}{M_s}\,\, u_R \, u_R \, d_R \,E_R\,\,,
 \end{align}
that lead to fast proton decay if not suppressed (for a similar analysis in the context of $SU(5)$ GUTs, see \cite{Kiritsis:2009sf}). The experimental bounds on the proton lifetime put an upper bound on the couplings constants $\kappa$ and $\kappa' $, which is given by \cite{Hinchliffe:1992ad}
 \begin{align}
 \kappa, \kappa' \leq 10^{-8} \,\,,
 \label{eq bounds on kappa}
\end{align} 
 where we used that $M_s \simeq 10^{18} \, GeV$. Note that, for this quiver, both quartic couplings are perturbatively forbidden, but will be induced by the instantons $E2_1$ and $E2_2$ and thus are suppressed by $\frac{e^{-S^{cl}_{E2_1}}}{M_s}$ and $\frac{e^{-S^{cl}_{E2_2}}}{M_s}$, respectively. 
 With $e^{-S^{cl}_{E2_1}} \simeq e^{-S^{cl}_{E2_2}}\simeq 10^{-2}$,  the suppression is not large enough to saturate the experimental bounds on the proton lifetime. Unless some symmetry of the internal compactification manifold prevents the presence of these dangerous dimension $5$ operators, or unless large worldsheet instanton suppressions significantly further suppress both operators, this setup suffers from rapid proton decay and has to be ruled out as  unrealistic.

Let us emphasize that the presence of these dangerous dimension $5$ operators is related to whether or not the right-handed down-quark $d_R$ is realized as antisymmetric of $SU(3)$. In case all of the down-quarks transform as bifundamental under $SU(3)$ and a $U(1)$ stack, both dimension 5 operators, $q_L\, q_L \,q_L \,L$ and  $u_R \, u_R \, d_R \,E_R$, are perturbatively absent. Furthermore, none of the instantons  required  to  generate one of the desired, but perturbatively forbidden couplings can induce one of these operators. However, in the case where the $d_R$ transform as antisymmetric under $SU(3)$, the operator  $u_R \, u_R \, d_R \,E_R$ could be perturbatively realized, depending on the transformation behavior of the other matter fields. In addition, instantons whose presence is required to induce missing couplings, might, in this case, generate the undesired couplings   $u_R \, u_R \, d_R \,E_R$ or $q_L\, q_L \,q_L \,L$ at the non-perturbative level. Whether or not the suppression factor of the coupling is sufficient to avoid rapid proton decay depends crucially on the suppression factor of the instanton, and thus is related to the scale of the perturbatively missing, but desired, Yukawa coupling generated by the same instanton.

To summarize, in this four-stack quiver all perturbatively missing Yukawa couplings can be generated without inducing any R-parity violating couplings or a $\mu$-term which is too large. Moreover, it allows for a mechanism which can account for the smallness of the neutrino masses. It also naturally gives rise to mass hierarchies similar to the ones observed in nature. A potential problem is the missing hierarchy between the $c$-quark mass and the $u$-quark mass, which cannot be explained via the two mechanisms discussed in chapter \ref{sec mass hierarchies}. On the other hand, world-sheet instantons can account for this hierarchy. A major flaw of this quiver is the presence of instanton induced dimension $5$ operators $q_L\, q_L \,q_L \,L$ and  $u_R \, u_R \, d_R \,E_R$, which generically are not suppressed enough to explain the large proton lifetime observed in experiments.

\section{Five-stack Quivers \label{sec five stack}}

In the previous section, we saw that none of the four-stack quivers give rise to the observed hierarchies for all the matter fields without inducing R-parity violating couplings or a too large $\mu$-term. We presented one setup which exhibits three different mass scales for the down-quarks and the electrons, but only two mass scales for the up-quarks. The hierarchy for up-quarks is due to different origins for different families. Thus, in order to have a third mass scale, it is natural to increase the number of D-brane stacks realizing the MSSM by adding another $U(1)$-brane. This allows for additional origins for the matter fields, in particular for the right-handed up-quarks, and therefore the quiver may exhibit the desired hierarchies. The most promising hypercharge embedding, as discussed in appendix \ref{hypercharge appendix}, is the extended Madrid embedding
\begin{align}
U(1)_Y=\frac{1}{6}\,U(1)_a+\frac{1}{2}\,U(1)_c+ \frac{1}{2}\,U(1)_d +\frac{1}{2}\,U(1)_e\,\,,
\label{eq: extended Madrid embedding}
\end{align}
which allows multiple potential origins for the MSSM matter fields. The possible origins are summarized below:
\begin{align*}
q_L\,:&\qquad (a,\ov{b}),\,\,\,\,(a,b)\\
u_R\,:&\qquad (\ov{a},\ov{c}),\,\,\,\,(\ov{a},\ov d),\,\,\,\,(\ov{a},\ov e)\\
d_R\,:&\qquad {\Yasymm}_a,\,\,\,\,(\ov{a},c),\,\,\,\,(\ov{a},d),\,\,\,\,(\ov{a},e)\\
L\,\,:&\qquad (b,\ov{c}),\,\,\,\,(\ov{b},\ov{c}),\,\,\,\,(b,\ov d),\,\,\,\,(\ov b,\ov d),\,\,\,\,(b,\ov e),\,\,\,\,(\ov b,\ov e)\\
E_R\,:&\qquad (c,d),\,\,\,\,(c,e),\,\,\,\,(d,e),\,\,\,\,{\Ysymm}_c,\,\,\,\,{\Ysymm}_d,\,\,\,\,{\Ysymm}_e\\
N_R\,:&\qquad {\Yasymm}_b,\,\,\,\,\ov{\Yasymm}_b,\,\,\,\,(c, \ov d),\,\,\,\, (\ov c, d),\,\,\,\,(c,\ov e),\,\,\,\, (\ov c, e),\,\,\,\,(d,\ov e),\,\,\,\, (\ov d, e)  \\
H_u\,:&\qquad (\ov b,c),\,\,\,\,(b,c),\,\,\,\,(\ov b,d),\,\,\,\,(b,d),\,\,\,\,(\ov b,e),\,\,\,\,(b,e)\\
H_d\,:&\qquad (b,\ov{c}),\,\,\,\,(\ov{b},\ov{c}),\,\,\,\,(b,\ov d),\,\,\,\,(\ov b,\ov d),\,\,\,\,(b,\ov e),\,\,\,\,(\ov b,\ov e)\,\,.
\end{align*}

As in the analysis performed in chapter \ref{sec four-stack}, we impose the criteria laid out in the systematic bottom-up analysis \cite{Cvetic:2009yh} and furthermore require that the D-brane quiver allows for the hierarchies observed in nature. We find that, even for five-stack realizations, there are only a few setups which allow for three different mass scales for the up-quarks, down-quarks and electrons, that are also compatible with experimental observations. Here we present two different types of setups. One gives rise to a Yukawa texture similar to the one seen in the four-stack quiver discussed in section \ref{sec four-stack}. However, a rigid $U(1)$ instanton \cite{Aganagic:2007py,GarciaEtxebarria:2007zv,Petersson:2007sc,GarciaEtxebarria:2008iw,Ferretti:2009tz} wrapping one of the three-cycles which is wrapped by one of the $U(1)$ branes induces an R-parity violating coupling, which rules out this setup as unrealistic. Moreover, this setup also suffers from the presence of dimension $5$ operators which lead to rapid proton decay.

In subsection \ref{sec perfect hierarchy}, we present  one of the few 5-stack quivers which not only exhibits three different mass scales for the up-quarks, down-quarks, and electrons and gives a natural explanation for the hierarchy between the $t$-quark mass and the masses of all other matter fields, but also overcomes the issue of a dangerously high proton decay rate encountered in the setups discussed in chapter \ref{sec four-stack} and section \ref{sec five stack U(1) instanton}. Thus this quiver passes all the stringent bottom-up constraints and provides a viable setup for further investigation.

\subsection{First Setup  -- R-Parity violation from $U(1)$ Instantons
 \label{sec five stack U(1) instanton}}

Table \ref{table spectrum setup with U(1) instanton} displays the origin and transformation behavior of all the MSSM matter fields for this setup, where the hypercharge is given by \eqref{eq: extended Madrid embedding} . We analyze it with respect to its Yukawa couplings  and textures. Since this analysis is very similar to the one performed in the previous chapter, we will be less detailed and instead we will just state the results.

\begin{table}[h] \centering
\begin{tabular}{|c|c|c|c|c|}
\hline
 Sector & Matter Fields &  Transformation & Multiplicity & Hypercharge\\
\hline \hline
 $ab$                            & $q_L$  & $(a,\overline{b})$ & $2$& $\frac{1}{6}$ \\
\hline
 $ab'$                            & $q_L$  & $(a,b)$ & $1$& $\frac{1}{6}$ \\
\hline
 $ac'$                            & $u_R$  & $(\overline{a},\overline{c})$  & $3 $ & $-\frac{2}{3}$ \\
\hline
$aa'$ & $d_R$ & ${\Yasymm}_a$ & $3$& $\frac{1}{3}$  \\
\hline
$bc$                            & $H_u$  & $(\overline{b},c)$  & $1$& $\frac{1}{2}$ \\
\hline
$bd$                            & $L$  & $(b,\overline{d})$ & $3$ & $ -\frac{1}{2} $  \\
\hline
$be'$                            & $H_d$  & $(\overline{b},\overline{e})$ & $1$ & $ -\frac{1}{2} $  \\
\hline
$bb'$                            & $N_R$  & $\overline{{\Yasymm}}_b $  & $1$ & $0$ \\
\hline
$cd$                            & $N_R$  & $(\ov c,d)$  & $1 $ & $0$ \\
\hline
$cd'$                            & $E_R$  & $(c,d)$  & $1 $ & $1$ \\
\hline
$ce$                            & $N_R$  & $(\ov c,e)$  & $1 $ & $0$ \\
\hline
$cc'$                            & $E_R$  & $\Ysymm_c$  & $1 $ & $1$ \\
\hline
$de'$                            & $E_R$  & $(d,e)$  & $1 $ & $1$ \\
\hline
\end{tabular}
\caption{\small {Spectrum for setup 1 with $U(1)_Y=\frac{1}{6}\,U(1)_a+\frac{1}{2}\,U(1)_c+ \frac{1}{2}\,U(1)_d +\frac{1}{2}\,U(1)_e$.}} 
\label{table spectrum setup with U(1) instanton}
\end{table}\vspace{5pt}

After taking into account non-perturbative effects, the up-quark Yukawa matrix takes a form similar to the one in the four-stack quiver discussed in the previous chapter
\begin{align*}
M^{IJ}_{U}= \,\langle \, H_u \, \rangle   \,\,\left(
\begin{array}{ccc}
g^u_{11} & g^u_{12}  & g^u_{13}\\
g^u_{21}\,e^{-S^{cl}_{E2_1}}  & g^u_{22}\,e^{-S^{cl}_{E2_1}} & g^u_{23} \,e^{-S^{cl}_{E2_1}}\\
g^u_{31} \,e^{-S^{cl}_{E2_1}} & g^u_{32} \,e^{-S^{cl}_{E2_1}} & g^u_{33} \,e^{-S^{cl}_{E2_1}}\\
\end{array}
\right)\,\,.
\end{align*}
Again, world-sheet instantons may account for the observed hierarchy between the two lightest families.

The down-quark mass matrix takes a slightly different form than in the four stack quiver discussed previously
\begin{align}
M^{IJ}_{D}= \,\langle \, H_d \, \rangle   \,\,\left(
\begin{array}{ccc}
a_{11} & a_{12} & a_{13} \\
a_{21} & a_{22} & a_{23} \\
a_{31} & a_{32} & a_{33} \\
\end{array}
\right)\,\,,
\label{eq: d-quark matrix 5-stack U(1)}
\end{align}
where the entries $a_{IJ}$ are given by
\begin{align*}
a_{IJ}=Y^{IJ}_D\,\, e^{-S^{cl}_{E2_2}}+Y'^{IJ}_D  \,\,e^{-S^{cl}_{E2_3}} +Y''^{IJ}_D\,\,  e^{-S^{cl}_{E2'_3}}\,\,.
\end{align*}
In contrast to the four-stack quiver, not all three instantons carry the same charge under the global $U(1)$ charges. The instanton $E2_2$ only populates the entries in the first row of the down-quark mass matrix \eqref{eq: d-quark matrix 5-stack U(1)}, thus $Y^{2J}_D=Y^{3J}_D=0$. The instantons $E2_3$ and $E2_3'$ give only non-vanishing contributions for the last two rows of \eqref{eq: d-quark matrix 5-stack U(1)}, thus $Y'^{1J}_D=Y''^{1J}_D=0$. Note that the latter two instantons give rise to factorizable Yukawa matrices. In order to get non-vanishing masses for all three families, the presence of both instantons, $E2_3$ and $E2'_3$,  is required. To match  experimental observations, the ratios of the suppression factors have to satisfy
\begin{align*}
m_b : m_s : m_d  \simeq  e^{-S^{cl}_{E2_2}} :  e^{-S^{cl}_{E2_3}}:  e^{-S^{cl}_{E2'_3}}\,\,,
\end{align*}
where $e^{-S^{cl}_{E2_2}}$ is expected to be in the range $10^{-2}-5\cdot 10^{-2}$.

Since all three right-handed electrons $E_R$ arise from different sectors, the global $U(1)$ charge carried by the Yukawa coupling is different for each family. The resulting electron Yukawa matrix takes the form
\begin{align*}
M^{IJ}_{E}= \,\langle \, H_d \, \rangle  \,\,\left(
\begin{array}{ccc}
g^e_{11} & g^e_{12}\,e^{-S^{cl}_{E2_4}} & g^e_{13}\,e^{-S^{cl}_{E2_5}}\\
g^e_{21} & g^e_{22}\,e^{-S^{cl}_{E2_4}}& g^e_{23}\,e^{-S^{cl}_{E2_5}} \\
g^e_{31} & g^e_{32}\,e^{-S^{cl}_{E2_4}} & g^e_{33}\,e^{-S^{cl}_{E2_5}} \\
\end{array}
\right)\,\,,
\end{align*}
where $g^e_{ij}$ denote the world-sheet instanton contributions. Note that in this quiver, in contrast to the four-stack quiver discussed in chapter \ref{sec four-stack}, the electron mass hierarchy between the three families arises from the fact that all electrons $E_R$ come from different sectors. Note, also, that this quiver predicts that the masses of the $t$-quark and the $\tau$-lepton are of the same order, which is in contradiction to experiments. Again, world-sheet instantons may suppress the $\tau$-lepton coupling in a way compatible with experimental observations.    

The Dirac neutrino mass matrix, after taking into account non-perturbative effects, takes the form
\begin{align*}
 M^{IJ}_{\nu}=\langle H_u \rangle\, \left(
 \begin{array}{ccc}
 g^{\nu}_{11} & g^{\nu}_{12} \, e^{-S_{E2_6}}  & g^{\nu}_{13} \, e^{-S_{E2_7}}\\
  g^{\nu}_{21} & g^{\nu}_{22} \, e^{-S_{E2_6}}  & g^{\nu}_{23} \, e^{-S_{E2_7}}\\
  g^{\nu}_{31} & g^{\nu}_{32} \, e^{-S_{E2_6}}  & g^{\nu}_{33} \, e^{-S_{E2_7}}\\
 \end{array}
\right)\,\,.
 \end{align*}
 Together with the Majorana mass matrix for the right-handed neutrinos
\begin{align*}
M^{IJ}_{N_R}= M_s \, \left( 
\begin{array}{ccc}
 g^{N_R}_{11}\, e^{-S_{E2_8}} & g^{N_R}_{12}\, e^{-S_{E2_9}}  & g^{N_R}_{13}\, e^{-S_{E2_{10}}} \\
 g^{N_R}_{12}\, e^{-S_{E2_9}} & g^{N_R}_{22}\, e^{-S_{E2_{11}}}  & g^{N_R}_{23}\, e^{-S_{E2_{12}}} \\
 g^{N_R}_{13}\, e^{-S_{E2_{10}}} & g^{N_R}_{23}\, e^{-S_{E2_{12}}}  & g^{N_R}_{33}\, e^{-S_{E2_{13}}} 
 \end{array}\right)\,\,,
\end{align*} 
where $M_s$ denotes the string mass, they can account for the smallness of the neutrino masses via the seesaw mechanism.
 
In contrast to the four-stack quiver, the $\mu$-term
\begin{align*}
<{H_u}_{(0,-1,1,0,0)} {H_d}_{(0,-1,0,0,-1)}>
\end{align*}
is perturbatively forbidden. It can be induced by an instanton with the intersection pattern 
\begin{align*}
I_{E2_{14}a}=0 \hspace{7mm} I_{E2_{14}b}=-1  \hspace{7mm} I_{E2_{14}c}=1   \hspace{7mm} I_{E2_{14}d}=0   \hspace{7mm}  I_{E2_{14}e}=-1,
\end{align*}
which can account for the order of the $\mu$-term, due to its non-perturbative nature.

As shown in \cite{Aganagic:2007py,GarciaEtxebarria:2007zv,Petersson:2007sc,GarciaEtxebarria:2008iw,Ferretti:2009tz}, a rigid $U(1)$ instanton wrapping the same cycle as one of the U(1)-branes or its orientifold images exhibits the right uncharged zero mode structure to contribute to the superpotential\footnote{Here we assume that the D-branes, and thus also the instanton, wrap rigid cycles in the internal manifold.}. In order to indeed contribute to the superpotential, such a $U(1)$ instanton also has to carry the right charged zero mode structure. Let us draw attention to the $U(1)$ instanton wrapping the same cycle as the D-brane $e$. Such an instanton carries the global $U(1)$ charges 
\begin{align}
Q_{E2}(a)=0 \,\,\,\,\,\,\,\, Q_{E2}(b)=-2\,\,\,\,\,\,\,\, Q_{E2}(c)=-1\,\,\,\,\,\,\,\, Q_{E2}(d)=1\,\,\, \,\,\,\,\, Q_{E2}(e)=0 
\end{align}
and thus can generate the R-parity violating coupling $L \,L \,E_R$. The suppression of this coupling depends on the volume of the three-cycle that the $U(1)$-brane $e$ wraps and therefore is related to the hypercharge coupling $g_Y$.

Due to the presence of the R-parity violating coupling  $L\, L\, E_R$, we rule out this setup as unrealistic. Note that, in contrast to the setups discussed in \cite{Cvetic:2009yh}, the instanton generating the R-parity violating coupling is not an instanton whose presence is required to induce one of the perturbatively missing, but desired, couplings. Rather, $U(1)$ instantons are present in every quiver and exhibit the right uncharged zero mode structure to induce superpotential terms. Whether or not they give contributions to the superpotential depends on the charged zero mode structure and thus on the spectrum charged under the $U(1)$-brane that wraps the same cycle as the instanton. 

It turns out that, for this quiver, a $U(1)$ instanton wrapping the same cycle as the D-brane $e$ induces an R-parity violating coupling. On the other hand, for different setups $U(1)$ instantons might generate one of the perturbatively missing, but desired, couplings.

The global embedding of D-instanton effects is still a very challenging task, mainly due to the fact that a generic instanton exhibits too many uncharged zero modes. For most examples of non-perturbative superpotential corrections in the literature, the generating instanton is an $O(1)$ instanton \cite{Argurio:2007qk,Argurio:2007vq,Bianchi:2007wy,Ibanez:2007rs}\footnote{As shown in \cite{Blumenhagen:2007bn,GarciaEtxebarria:2007zv,Cvetic:2008ws,GarciaEtxebarria:2008pi}, multi-instanton configurations can also contribute to the superpotential.}. Such an instanton wraps an orientifold invariant cycle and the undesired zero modes are projected out. However, for the known Calabi-Yau threefolds there is a limited number of such cycles, and they must additionally exhibit the right intersection pattern with the D-branes in order to contribute to the superpotential\footnote{See \cite{Cvetic:2007qj} for a globally consistent type I realization where D-instantons  generate Majorana mass terms for the right-handed neutrinos and a Polonyi-type 
superpotential term.}. Thus, having another source of D-instanton contributions, such as a rigid $U(1)$ instanton, may be helpful in the quest for a global string realization of the MSSM.  

While this setup does not exhibit the dangerous dimension $5$ operator  $q_L q_L q_L L$ at either the perturbative or non-perturbative level, the other dangerous dimension $5$ operator $u_R \, u_R \,d_R \,E_R$ is perturbatively realized and therefore generically does not satisfy the upper bound \eqref{eq bounds on kappa}. Thus, even apart from the $U(1)$ instanton induced R-parity violating coupling $L\, L\, E_R$, this quiver suffers from  rapid proton decay triggered by the quartic coupling $u_R \, u_R \,d_R \,E_R$.

Summarizing, this five-stack quiver exhibits a phenomenology very similar to that of the four-stack quiver discussed in chapter \ref{sec four-stack}. In particular, it suggests that the masses of the $c$-quark and the $u$-quark are of the same order and predicts that the masses of the $t$-quark and the $\tau$-lepton are of the same order.  Again, in principle, world-sheet instantons can account for the suppression of the $u$-quark and $\tau$-lepton mass. More serious flaws are that this quiver suffers from the presence of the R-parity violating coupling $L\, L\, E_R$, which is induced by a $U(1)$ instanton wrapping the same cycle as the D-brane $e$, and the perturbative presence of the dangerous dimension $5$ operator  $u_R \, u_R \,d_R \,E_R$, which leads to a disastrous proton decay rate.

\subsection{Second setup -- Realistic Yukawa textures\label{sec perfect hierarchy}}
Finally, let us discuss one of the few 5-stack quivers which not only naturally exhibits three different mass scales for electrons, up- and down-quarks and also naturally explains the hierarchy between the $t$-quark mass and the masses of all other MSSM matter fields, but also overcomes the serious issue of the dangerous dimension $5$ operators which lead to  rapid proton decay. Table \ref{table spectrum for 5-stack quiver setup 3} displays the origin and transformation behavior for all the MSSM matter fields, where again the hypercharge is given by the extended Madrid embedding \eqref{eq: extended Madrid embedding}.
\begin{table}[h] \centering
\begin{tabular}{|c|c|c|c|c|}
\hline
 Sector & Matter Fields &  Transformation & Multiplicity & Hypercharge\\
\hline \hline
  $ab$                            & $q_L$  & $(a,\overline{b})$ & $1$& $\frac{1}{6}$ \\
 \hline
 $ab'$                            & $q_L$  & $(a,b)$ & $2$& $\frac{1}{6}$ \\
\hline
 $ac'$                            & $u_R$  & $(\overline{a},\overline{c})$  & $2 $ & $-\frac{2}{3}$ \\
\hline
 $ad'$                            & $u_R$  & $(\overline{a}, \ov d)$  & $1 $ & $-\frac{2}{3}$ \\
\hline
$aa'$ & $d_R$ & ${\Yasymm}_a$ & $3$& $\frac{1}{3}$  \\
\hline
$bc'$                            & $H_u$  & $(b,c)$ & $1$ & $\frac{1}{2}  $  \\
\hline
$bd'$                            & $L$  & $(\ov b,\overline{d})$  & $3$& $-\frac{1}{2}$ \\
\hline
$be'$                            & $H_d$  & $(\overline{b},\overline{e})$ & $1$ & $\frac{1}{2}  $  \\
\hline
$ce'$                            & $E_R$  & $(c,e)$  & $2 $ & $1$ \\
\hline
$ce$                            & $N_R$  & $(\ov c, e)$  & $1 $ & $0$ \\
\hline
$dd'$                            & $E_R$  & $\Ysymm_d$  & $1 $ & $1$ \\
\hline
$de$                            & $N_R$  & $(  \ov d, e)$  & $2 $ & $0$ \\
\hline
\end{tabular}
\caption{\small {Spectrum for setup 2 with $U(1)_Y=\frac{1}{6}\,U(1)_a+\frac{1}{2}\,U(1)_c+ \frac{1}{2}\,U(1)_d +\frac{1}{2}\,U(1)_e$.} }
\label{table spectrum for 5-stack quiver setup 3}
\end{table}

After taking into account non-perturbative effects, the up-quark Yukawa coupling matrix takes the form
\begin{align}
M^{IJ}_{U}= \,\langle \, H_u \, \rangle   \,\,\left(
\begin{array}{ccc}
g^u_{11} & g^u_{12}  & 0\\
g^u_{21}\,e^{-S^{cl}_{E2_1}}& g^u_{22} \,e^{-S^{cl}_{E2_1}}& g^u_{23} \,e^{-S^{cl}_{E2_2}}\\
g^u_{31}\,e^{-S^{cl}_{E2_1}}& g^u_{32}\,e^{-S^{cl}_{E2_1}}& g^u_{33} \,e^{-S^{cl}_{E2_2}}\\
\end{array}
\right)\,\,,
\label{eq: up-quark matrix perfect setup }
\end{align}
where the $g^u_{ij}$, as before, denote the world-sheet instanton contributions. Note that the $t$-quark coupling is realized perturbatively and the up-quark matrix \eqref{eq: up-quark matrix perfect setup } generically exhibits three different mass scales for the up-quarks. In order to match the observed hierarchies, the ratios of the suppression factors are expected to be
\begin{align*}
m_t : m_c : m_u \simeq 1:e^{-S^{cl}_{E2_1}}: e^{-S^{cl}_{E2_2}} \,\,.
\end{align*}

The down-quark mass matrix takes the same form as for the quiver investigated in section \ref{sec five stack U(1) instanton} (see eq. \eqref{eq: d-quark matrix 5-stack U(1)}). To match  experimental observations, we expect the ratios of the suppression factors to satisfy\footnote{Note that, for this setup, we substituted $E2_2$ by $E2_3$, $E2_3$ by $E2_4$ and $E2'_3$ by $E2'_4$ compared to the quiver in section \ref{sec five stack U(1) instanton}.}
\begin{align*}
m_b : m_s : m_d  \simeq  e^{-S^{cl}_{E2_3}} :  e^{-S^{cl}_{E2_4}}:  e^{-S^{cl}_{E2'_4}}\,\,,
\end{align*}
where $e^{-S^{cl}_{E2_3}}$ is expected to be in the range $10^{-2}-5\cdot 10^{-2}$.

The electron Yukawa coupling for this setup is of the form
\begin{align}
M^{IJ}_{E}= \, \langle \,H_d\,\rangle  \,\,\left(
\begin{array}{ccc}
b_{11} & b_{12} & b_{13} \\
b_{21} & b_{22} & b_{23} \\
b_{31} & b_{32} & b_{33} \\
\end{array}
\right)\,\,,
\label{eq: electron matrix 5-stack perfect}
\end{align}
where the entries take a similar form as for the down-quark coupling
\begin{align*}
b_{IJ}=Y^{IJ}_E\,\, e^{-S^{cl}_{E2_5}}+Y'^{IJ}_E  \,\,e^{-S^{cl}_{E2_6}} +Y''^{IJ}_E\,\,  e^{-S^{cl}_{E2'_6}}\,\,,
\end{align*}
with $Y^{J2}_E=Y^{J3}_E=0$ and $Y'^{J1}_E=Y''^{J1}_E=0$. The instanton $E2_6$ induces a factorizable Yukawa matrix, and thus a second instanton $E2'_6$ with the same intersection pattern as $E2_6$ is needed, together with the instanton $E2_5$, to give masses to all three families. Note that, in contrast to the setup discussed in section \ref{sec five stack U(1) instanton} , the $\tau$-lepton mass is perturbatively absent. To match experimental observations, we expect the suppression factors to satisfy
\begin{align*}
m_{\tau}: m_{\mu}:m_{e}  \simeq e^{-S^{cl}_{E2_5}} :e^{-S^{cl}_{E2_6}} : e^{-S^{cl}_{E2_6'}} \simeq 10^{-2} : 10^{-4} : 10^{-7}\,\,.
\end{align*}

Due to the absence of perturbative Dirac masses, the smallness of the neutrino masses can be explained by the non-perturbative generation of the Dirac masses, where the instanton suppression factor is in the range $10^{-14}-10^{-11}$ \cite{Cvetic:2008hi}. The  Dirac mass matrix takes the form
\begin{align*}
M^{IJ}_{\nu}= \, \langle \,H_u\,\rangle  \,\,\left(
\begin{array}{ccc}
g^{\nu}_{11} \,e^{-S^{cl}_{E2_8}}& g^{\nu}_{12}\,e^{-S^{cl}_{E2_9}} + g'^{\nu}_{12}\,e^{-S^{cl}_{E2'_9}}  & g^{\nu}_{13} \,e^{-S^{cl}_{E2_9}}  +g'^{\nu}_{13}\,e^{-S^{cl}_{E2'_9}}  \\
g^{\nu}_{21}\,e^{-S^{cl}_{E2_8}}& g^{\nu}_{22} \,e^{-S^{cl}_{E2_9}} + g'^{\nu}_{22}\,e^{-S^{cl}_{E2'_9}} & g^{\nu}_{23} \,e^{-S^{cl}_{E2_9}} + g'^{\nu}_{23}\,e^{-S^{cl}_{E2'_9}} \\
g^{\nu}_{31}\,e^{-S^{cl}_{E2_8}} & g^{\nu}_{32} \,e^{-S^{cl}_{E2_9}} +g'^{\nu}_{32}\,e^{-S^{cl}_{E2'_9}}  & g^{\nu}_{33} \,e^{-S^{cl}_{E2_9}} + g'^{\nu}_{33}\,e^{-S^{cl}_{E2'_9}} \\
\end{array}
\right)\,\,.
\end{align*}
The instantons $E2_9$ and $E2'_9$ induce a factorizable Yukawa matrix, thus the presence of both instantons is required to induce masses for all the neutrinos\footnote{Note that experiments do not yet rule out the possibility of one neutrino family being massless. If we allow for a massless neutrino family, only the presence of two instantons of $E2_8$,  $E2_9$  and $E2'_9$ is required.}.
The $\mu$ term is perturbatively forbidden and can be generated by a D-instanton with the intersection pattern
\begin{align*}
I_{E_{10}a}=0 \qquad I_{E2_{10}b}=0 \qquad I_{E2_{10}c}=1 \qquad I_{E2_{10}d}=0 \qquad I_{E2_{10}e}=-1
\end{align*}
To get a $\mu$-term of the order $100 \,GeV$, the suppression factor of $E2_{10}$ is expected to be of the order $10^{-16}$.

Note that, in contrast to the setup discussed in chapter \ref{sec four-stack}, this setup does not exhibit the dangerous dimension five operator $q_L\,q_L\,q_L\,L$ at either the perturbative or non-perturbative level. On the other hand, the quartic coupling $u_R \, u_R\, d_R \,E_R$ is induced by the same instanton that generates Dirac mass term for neutrinos, $E2_8$, which exhibits the intersection pattern
\begin{align*}
 I_{E2_8a}=0 \qquad  I_{E2_8b}=0  \qquad I_{E2_8c}=0 \qquad  I_{E2_8d}=-1 \qquad I_{E2_8e}=1\qquad I^{{\cal N}=2}_{E2_8c}=1\,\,.
\end{align*}
Note, however, that in order to explain the small observed Neutrino masses the suppression factor of $E2_8$ is expected to be in the range $10^{-14}-10^{-11}$, which is enough suppression to saturate the present bounds on the proton lifetime (see eq. \eqref{eq bounds on kappa}). Moreover, since experiments do not exclude the possibility of one massless neutrino, the presence of the instanton $E2_8$ is not required. In that case, the dimension $5$ operator  $u_R \, u_R\, d_R \,E_R$ is not even induced. We conclude that, in contrast to the previous examples discussed in chapter \ref{sec four-stack} and section \ref{sec five stack U(1) instanton}, this quiver does not suffer from a large proton decay rate, and thus provides a viable setup which gives rise to realistic phenomenology.   

Let us summarize the features of this quiver. It naturally exhibits three different mass scales for the up-quarks, down-quarks and electrons, as observed in nature, and naturally explains the hierarchy between the $t$-quark mass and the masses of all other matter fields. It allows a natural explanation for the neutrino and Higgs mass scales, due to their non-perturbative nature. Moreover, it overcomes the serious issue of a disastrous proton decay rate. Therefore it provides a viable setup for further phenomenological investigation.

\section{Conclusion}

In this work, we discuss the potential origin of MSSM mass hierarchies in orientifold string compactifications. Such compactifications generically exhibit a large class of global $U(1)$'s, which are remnants of the generalized Green-Schwarz mechanism. These global $U(1)$'s must be respected by perturbative string theory and thus often times forbid desired couplings. The missing couplings can be generated via higher-order couplings containing the VEV's of SM singlets or by D-instantons, both of which lead naturally to a suppression of the induced coupling relative to perturbatively realized interactions, thus potentially giving rise to interesting Yukawa textures.

In chapter \ref{sec mass hierarchies}, we present two different methods to obtain mass hierarchies between different families of the MSSM fields. One potential origin is that different families arise from different sectors\cite{Anastasopoulos:2009mr}. Then, for each family, a different instanton or singlet is required to induce the perturbatively missing coupling, which leads to different suppression factors for different generations. The other potential mechanism is purely non-perturbative and relies on the fact that sometimes an instanton-induced Yukawa matrix factorizes, in which case it only gives rise to one massive family. Thus, three different instantons carrying the same charge under the global $U(1)$'s are required to give masses to all three families. These three instantons generically wrap different cycles in the internal manifold and therefore exhibit different suppression  factors. The resulting mass matrix then gives three different masses, which are of the order of the three suppression factors of the instantons.

Any perturbative coupling is suppressed by world-sheet instantons, which in principle can account for the observed mass hierarchies, as shown in 
\cite{Aldazabal:2000cn, Cvetic:2001tj, Cvetic:2002wh,Cremades:2003qj,Cvetic:2003ch}. However, these world-sheet instantons depend crucially on the geometry of the internal manifold, as well as on the open string moduli, and therefore a large amount of fine-tuning is required to obtain the hierarchies observed in nature. Another potential source of mass hierarchies in orientifold compactifications relies on some symmetries of the internal manifold, which leads to factorizable Yukawa matrices on the perturbative level. For such compactifications, only one family becomes massive, while the others remain massless and receive masses via radiative corrections \cite{Abel:2003yh}. However, such factorization depends on the geometry of the internal manifold and does not appear for a generic compactification manifold.

In \cite{Cvetic:2009yh}, the authors systematically analyzed four-stack quivers which give rise to the exact MSSM spectrum and allow for the generation of the MSSM superpotential, perturbatively or non-perturbatively, without inducing any phenomenological drawbacks, such as R-parity violating couplings. In addition, they require that the quivers provide a mechanism which explains the smallness of the neutrino masses. This analysis was performed independently of a concrete global realization and therefore the results are independent of the geometrical specifics of the internal space.  In this work, we extend this analysis by requiring that the D-brane quivers must also allow for a natural explanation of the observed hierarchies, where the origin of the hierarchies is due to the two mechanisms presented in chapter \ref{sec mass hierarchies}.

We find that only one four-stack quiver exhibits a semi-realistic Yukawa texture. In chapter \ref{sec four-stack}, we analyze this  quiver with respect to its Yukawa structure. We show that D-instantons can account for the generation of perturbatively missing, but desired, couplings without inducing R-parity violating couplings or a too large $\mu$-term. Moreover, we show that this quiver naturally allows for three different mass scales for the down-quarks and the electrons, and also explains the hierarchy between the $t$-quark mass and the masses of all other MSSM matter fields. However, it predicts that the $c$-quark and $u$-quark masses are of the same order. In principle, world-sheet instantons can account for the hierarchy between the two lightest families of the up-quarks, but this often involves some amount of fine-tuning.

In chapter \ref{sec five stack}, we allow for an additional $U(1)$-brane stack and, as expected, we find many more setups which exhibit a Yukawa texture similar to the one encountered in chapter \ref{sec four-stack} for the four-stack quiver.  A subclass of those D-brane quivers allows for a $U(1)$ instanton, which wraps the same cycle as one of the $U(1)$ branes, whose global $U(1)$ charge and uncharged zero mode structure is such that it generates a superpotential contribution. 
Let us emphasize that if a $U(1)$ instanton exhibits the proper intersection pattern, and thus the proper global $U(1)$ charge, to induce a superpotential coupling, then the presence of the latter is guaranteed, independent of the concrete global embedding. In section \ref{sec five stack U(1) instanton} we present such a quiver, where a $U(1)$ instanton induces an R-parity violating coupling, and thus we rule out the setup as unrealistic. In a future systematic bottom-up analysis, similar to the ones performed in \cite{Cvetic:2009yh}, it would be interesting to require the absence of $U(1)$ instantons which generate R-parity violating couplings. 

Whether or not a $U(1)$ instanton can generate a particular coupling depends crucially on the spectrum charged under the $U(1)$-brane that the instanton wraps. Though in the example discussed in section \ref{sec five stack U(1) instanton} it induces an undesired R-parity violating coupling, for different setups it may generate a perturbatively forbidden, but desired, Yukawa coupling. Given how difficult it is to find global embeddings of D-instanton effects, we believe that $U(1)$ instantons are a useful additional tool in generating some of the desired, but missing, MSSM superpotential couplings.

Both the four-stack quiver and the five-stack quiver, discussed in chapter \ref{sec four-stack} and section \ref{sec five stack U(1) instanton}, respectively, suffer from the presence of quartic couplings, whose presence gives rise to  rapid proton decay, which is not compatible with experimental observations (see \cite{Kiritsis:2009sf} for a similar discussion in the context of $SU(5)$ GUT orientifolds). The presence of of such dimension $5$ operators is related to the fact that the right-handed down-quarks are realized as antisymmetrics of $SU(3)$, which may explain the different mass scales for the three down-quarks, via the mechanism described in section \ref{sec factorizable Yukawa matrices}. Thus, there is a tension between obtaining the observed mass hierarchies of the MSSM and avoiding a disastrous proton decay rate.

In section \ref{sec perfect hierarchy}, we present one of the few five-stack quivers that passes all of the stringent bottom-up constraints without fine-tuning.
It gives rise to three different mass scales for the up-, down-quarks and electrons, as observed in nature, and naturally explains the hierarchy between the $t$-quark mass and the masses of all other matter fields. It also gives an explanation for the small neutrino masses and the $\mu$-term, due to their non-perturbative nature. Moreover, the dangerous dimension $5$ operators that could lead to  rapid proton decay are either absent or sufficiently suppressed. Thus, this quiver is a promising starting point for further detailed phenomenological investigation.

A generic pattern of MSSM orientifold compactifications is that different families tend to arise from the same sector, even though there are many potential origins they can arise from. In such cases the mechanism discussed in section \ref{sec Family splitting} cannot be applied to get the observed mass hierarchies. One might entertain the idea of allowing for additional chiral singlets, which acquire a VEV and may induce perturbatively missing couplings via higher order couplings. For such setups, the presence of the additional chiral singlets will have dramatic effects on the constraints arising from tadpole cancellation and the presence of a massless hypercharge $U(1)_Y$. In such a case, it would be interesting to see if family splitting is still phenomenologically disfavored. 

\vspace{2.5cm}
{\bf Acknowledgments}\\
We thank M. Ambroso, M. Bianchi, T. Brelidze, B. G. Chen, I. Garc\'ia-Etxebarria, L. E. Ib{\'a}{\~n}ez, E. Kiritsis, A. Lionetto, B. Schellekens and T. Weigand for
useful discussions. The work is supported by the DOE  Grant DOE-EY-76-02-3071, the NSF RTG grant DMS-0636606 and the Fay R. and Eugene L. Langberg
Chair.

\newpage

\appendix
\section{Top-down and bottom-up constraints
\label{app top-down bottom-up}}
In this appendix, we briefly summarize the constraints we require D-brane quivers to satisfy. For a more detailed description, we refer the reader to \cite{Cvetic:2009yh}. Let us distinguish between the two different classes of constraints, \emph{top-down} and \emph{bottom-up} constraints. The former include constraints on the chiral matter field transformation behavior arising from tadpole cancellation and from the presence of a massless hypercharge $U(1)_Y$.
The bottom-up constraints are due to experimental observations and include the absence of R-parity violating couplings on the perturbative and non-perturbative level, as well as the presence of a mechanism which explains the smallness of the neutrino masses.

\begin{itemize}
\item[$\bullet$] All the MSSM matter content and the right-handed neutrino, apart from the Higgs
fields, appear as chiral fields at intersections between two D-brane stacks of the respective quiver. That implies 
that all of the MSSM matter fields are charged only under the four or five D-brane gauge groups. Moreover, we require the absence of additional chiral fields charged under the gauge groups of the four or five D-branes.
\item[$\bullet$] As discussed in \cite{Cvetic:2009yh}, tadpole cancellation, which is a condition on the cycles that the D-branes wrap, imposes constraints on the transformation behavior of the chiral matter. For a stack of $N_a$ D-branes  with $N_a>1$, the constraints read
\begin{align}
\#(a) - \#({\ov a})  + (N_a-4)\#( \, \Yasymm_a) + (N_a+4) \#
(\Ysymm_a)=0 \,\,,\label{eq constraint1}
\end{align}
while for $N_a=1$ it is slightly modified and takes the form
\begin{align}
\#(a) - \#({\ov a}) + 5 \# (\Ysymm_a)=0  \qquad \text{mod} \,3\,\,.
\label{eq constraint2}
\end{align}
We require the constraints to be satisfied by the MSSM matter content for all D-brane stacks.
\item[$\bullet$] In a fashion similar to tadpole cancellation, the presence of a massless $U(1)_Y$ puts constraints on the cycles that the D-branes wrap. These again imply constraints on the transformation behaviour of the chiral matter, which are given by
\begin{align}
 \sum_{x \neq a} q_x\,
N_x \#(a,{\ov x}) - \sum_{x \neq a} q_x\, N_x \#(a,x) = q_a\,N_a \,\Big(\#(\Ysymm_a) + \# (\, \Yasymm_a)\Big) \,\,
\label{eq massless constraint non-abelian}
\end{align}

for $N_a>1$ and
\begin{align}\sum_{x \neq a} q_x\, N_x \#(a,{\ov x}) -
\sum_{x \neq a} q_x\, N_x \#(a,x)= q_a\,\frac{\#(a) - \#({\ov a}) + 8 \#
(\Ysymm_a)}{3}  
\label{eq massless constraint abelian}
\end{align}
for $N_a=1$.
\item[$\bullet$]All the MSSM Yukawa couplings for all three families are realized, either perturbatively, non-perturbatively or via higher order couplings. In the latter case the MSSM singlets are non-chiral and acquire a VEV by brane splitting. 
\item[$\bullet$] We forbid any R-parity violating couplings on the perturbative and non-perturbative level.
\item[$\bullet$] Often times an instanton which is required to generate the Yukawa couplings
also induces a tadpole $N_R$ and thus an instability for the setup. We rule out any setup which requires the presence of such an instanton.
\item[$\bullet$]
We rule out setups which lead to a large family mixing in the quark Yukawa couplings \cite{Ibanez:2008my,Leontaris:2009ci,Cvetic:2009yh}.
\item[$\bullet$]  The D-brane quiver must allow for a mechanism which gives a $\mu$-term of the observed order.
\item[$\bullet$] We require that the D-brane quiver exhibits a mechanism which accounts for the smallness of the neutrino masses.
\item[$\bullet$] We require that the $t$-quark Yukawa coupling is realized perturbatively.
\end{itemize}

\section{Neutrino mass matrix
\label{app neutrino mass}}

Here we investigate whether or not the Majorana mass matrix encountered in section \ref{sec four-stack} factorizes. If it does, then only one family acquires a seesaw mass, while for the other two families the mass is roughly given by the Dirac mass term. The intersection pattern \eqref{eq: intersection pattern E2_4}  suggests that the instanton exhibits four charged zero modes $ \lambda^{\alpha}_b$, where $\alpha$ denotes the family index, $\alpha\, \epsilon \, (1,2)$. The instanton action generically takes form
\begin{align}
S_{E2}= S^{cl}_{E2}+ Y^{I}_{\alpha \beta} \,\,   \lambda^{\alpha}_b \, N^I_R\,  \lambda^{\beta}_b\,\,,
\label{eq: instanton action nutrino masses}
\end{align}
where $Y^I_{\alpha \beta}$ denotes the world-sheet instantons and generically depends on the zero mode family indices $\alpha$ and $\beta$, as well as on the neutrino family index $I$. Performing the path integral over the charged zero modes
\begin{align}
\int  d^4 x\, d^2 \theta \,d^4  \lambda_b\, e^{-S_{E2}} 
\end{align}
gives the Majorana mass matrix for the right-handed neutrinos, which takes the form
\begin{align}
M^{IJ}_{N_R}= e^{-S^{cl}_{E2}} \,\,M_s \,\,\left(
\begin{array}{ccc}
M^{\nu}_{11} & M^{\nu}_{12}  & M^{\nu}_{13} \,\\
M^{\nu}_{21}& M^{\nu}_{22} & M^{\nu}_{23} \\
M^{\nu}_{31} & M^{\nu}_{32} & M^{\nu}_{33} \\
\end{array}
\right)\,\,,
\label{eq: Majorana mass matrix factorizable}
\end{align}
where the entries take the form
\begin{align}
M^{\nu}_{IJ}= 2\,Y^{I}_{11} \, Y^{J}_{22} - Y^{I}_{12}\, Y^{J}_{12} - Y^I_{12}\, Y^J_{21} - Y^{I}_{21} \, Y^{J}_{12} - Y^{I}_{21}\, Y^{J}_{21} +2\, Y^I_{22}\, Y^J_{11} \,\,.
\label{eq: majorana mass entries}
\end{align}
Thus, generically, the Majorana mass matrix does not factorize, and for all three families a Majorana mass term of the order $M_s \,e^{-S^{cl}_{E2}}$ is induced.\\

\section{Discussion of Five-Stack Hypercharges}
\label{hypercharge appendix}
There are multiple five-stack hypercharge embeddings possibly consistent with the exact MSSM spectrum and tadpole constraints. Many of these hypercharges, however, can be ruled out by requiring that the quiver exhibits three different mass scales for the up-quarks, down-quarks and leptons. Here we allow for the two different mechanisms discussed in chapter \ref{sec mass hierarchies} to give rise to such hierarchies. Only if the quarks are realized as antisymmetric of $SU(3)$ do non-perturbative effects give rise to a factorizable Yukawa matrix, and thus to interesting mass hierarchies. Otherwise the induced Yukawa matrix does not factorize, and desired hierarchies must have their origin in the fact that different families arise from different sectors. In this appendix we analyze which of the five-stack hypercharge embeddings simultaneously give rise to three hierarchies and satisfy the constraints laid out in appendix \ref{app top-down bottom-up}.

For five-stack quivers, the hypercharges take one of two forms once we restrict the spectrum to the exact MSSM and impose the top-down constraints arising from tadpole cancellation. All such hypercharges take one of the following forms

\begin{align} \label{eq: hypchoice 1}
U(1)_Y&=-\frac{1}{3}U(1)_a - \frac{1}{2}U(1)_b+ \sum_{i} q_i U(1)_i\\ 
U(1)_Y&=\frac{1}{6}U(1)_a+\frac{1}{2}U(1)_c+ \sum_{i} q_i U(1)_i\,\,,
\label{eq: hypchoice 2}
\end{align}
for some set of $q_i$'s, where the sum contains the contributions from the $U(1)$ brane stacks\footnote{Since models with a $U(1)$ brane $i$ have a symmetry under a simultaneous swapping of $q_i \leftrightarrow  -q_i$, transformation behavior $i\leftrightarrow \ov{i}$ and branes $i\leftrightarrow i'$, we fix $q_i\ge0$ for all $i$.}.

For the first possible hypercharge form \eqref{eq: hypchoice 1}, $d_R$ may transform only as $(\ov{a},i)$ or $(\ov{a},\ov{i})$, where $i$ is a $U(1)$ stack with $q_i=0$. In the case where no $q_i=1$, $u_R$ must transform as $\Yasymm_a$. In this case, the $d_R$ can transform as $(\ov{a},i)$ or $(\ov{a},\ov{i})$, but not both, since then the R-parity violating coupling $u_R\,d_R\,d_R$ would be perturbatively allowed. This implies that all three U(1) branes must have $q_i=0$ in order to give rise to three mass hierarchies for the down-quarks. Thus, the only allowed hypercharge with no $q_i=1$ is
\begin{align*}
U(1)_Y=-\frac{1}{3}U(1)_a-\frac{1}{2}U(1)_b\,\,.
\end{align*}
However, having three different origins for $d_R$ requires instantons to generate $q^I_L\, H_d\, d^J_R$ for all three families. These same instantons will also generate the R-parity violating terms  $u^I_R \, d^J_R\, d^K_R$ for some choice of $I$, $J$ and $K$, and thus this case also fails.

In the case where $q_i=1$ for a $U(1)$ stack, the existence of three down-quark mass hierarchies requires at least two $U(1)$ branes with $q_i=0$, and thus
\begin{align*}
U(1)_Y=-\frac{1}{3}U(1)_a-\frac{1}{2}U(1)_b+U(1)_e\,\,.
\end{align*}
Here all $u_R$'s must transform as $(\ov{a},\ov{e})$ to prevent $u^I_R\,d^J_R\, d^K_R$ from being perturbatively allowed, which gives no mass hierarchy between the up-quark families. Thus there are no D-brane quivers with the hypercharge embedding of type \eqref{eq: hypchoice 1} which give rise to the observed mass hierarchies without inducing R-parity violating couplings.

For the second possible hypercharge  \eqref{eq: hypchoice 2}, the $d_R$'s can transform as antisymmetric $\Yasymm_a$ of $SU(3)$, and thus in principle there is, independent of the choices of $q_i$ in \eqref{eq: hypchoice 2}, a way to get three different mass scales for the down-quarks. The $u_R$'s transform only as $(\ov{a},\ov{i})$ for a $U(1)$ brane $i$ with $q_i=\frac{1}{2}$. On the other hand the $q_L$'s can have two different transformation behaviors,  as $(a,\ov{b})$ or $(a,b)$. To obtain one of the three Yukawa matrices displayed in equation \eqref{eq: Yukawa matrices quarks}, the choice of hypercharges must allow for at least two different origins for the $u_R$'s, and thus at least two of the $q_i$'s in \eqref{eq: hypchoice 2} take the value $\frac{1}{2}$. Since we require that all D-branes are populated by at least one of the MSSM fields, there are only three different choices of hypercharge which could potentially give rise to three different mass scales for the up-quarks, down-quarks and electrons, while also being compatible with the constraints arising from tadpole cancellation. They are displayed below
\begin{align*}
U(1)_Y&=\frac{1}{6}U(1)_a+\frac{1}{2}U(1)_c+\frac{1}{2}U(1)_d\\
U(1)_Y&=\frac{1}{6}U(1)_a+\frac{1}{2}U(1)_c+\frac{1}{2}U(1)_d+\frac{3}{2}U(1)_e\\
U(1)_Y&=\frac{1}{6}U(1)_a+\frac{1}{2}U(1)_c+\frac{1}{2}U(1)_d+\frac{1}{2}U(1)_e\,\,.
\end{align*}
For these three hypercharge choices, we imposed the top-down and bottom-up constraints presented in appendix \ref{app top-down bottom-up} and analyzed which of of the D-brane quivers exhibit three different mass scales for the up-quarks, down-quarks and electrons. All solutions for the first two choices of hypercharge which satisfy the constraints do not have the phenomenologically desired hierarchies. For the last hypercharge, the extended Madrid embedding, we do find some realizations which do not suffer from this drawback, and two examples are discussed in chapter \ref{sec five stack}.

\clearpage \nocite{*}
\bibliography{rev}
\bibliographystyle{utphys}

 \end{document}